\documentclass[journal]{IEEEtran}
\usepackage{cite}
\usepackage[cmex10]{amsmath}
\usepackage{amssymb}
\usepackage{mathtools}
\usepackage{graphicx}
\usepackage{multirow}
\usepackage{booktabs}
\usepackage{color}
\usepackage{algorithm}
\usepackage{algorithmicx,algpseudocode}
\usepackage{url}
\usepackage{float}
\usepackage{textcomp}

\begin{document}

\title{\huge Storage and Transmission Capacity Requirements \\ of a Remote Solar Power Generation System}

\author{Yue Chen, Wei Wei, Cheng Wang, Miadreza Shafie-khah, Jo{\~a}o P. S. Catal{\~a}o

\thanks{This work has been submitted to the IEEE for possible publication.
Copyright may be transferred without notice, after which this version may
no longer be accessible.}
\thanks{Y. Chen is with the Department of Mechanical and Automation Engineering, the Chinese University of Hong Kong, Hong Kong, China. (e-mail: yuechen@mae.cuhk.edu.hk).}
\thanks{W. Wei is with the State Key Laboratory of Power Systems, Department of Electrical Engineering, Tsinghua University, Beijing 100084, China. (e-mail: wei-wei04@mails.tsinghua.edu.cn).}
\thanks{C. Wang is with the State Key Laboratory of Alternate Electrical Power System with Renewable Energy Sources, North China Electric Power University, Beijing 102206, China (e-mail: chengwang@ncepu.edu.cn).}
\thanks{M. Shafie-khah is with School of Technology and Innovations, University of Vaasa, 65200 Vaasa, Finland (e-mail: miadreza@gmail.com).}
\thanks{J. P. S. Catal{\~a}o is with Faculty of Engineering of the University of Porto and INESC TEC, Porto 4200-465, Portugal (e-mail: catalao@fe.up.pt).}
}

\maketitle

\begin{abstract}
Large solar power stations usually locate in remote areas and connect to the main grid via a long transmission line. Energy storage unit is deployed locally with the solar plant to smooth its output. Capacities of the grid-connection transmission line and the energy storage unit have a significant impact on the utilization rate of solar energy, as well as the investment cost. This paper characterizes the feasible set of capacity parameters under a given solar spillage rate and a fixed investment budget. A linear programming based projection algorithm is proposed to obtain such a feasible set, offering valuable references for system planning and policy making.
\end{abstract}


\section{Introduction}

\IEEEPARstart{T}{he} penetration of wind and solar generation in power systems has witnessed dramatic growth during the past decade. However, the solar energy is intermittent; no power can be produced during the night, calling for  sufficient backup capacity to mitigate the intra-hour$/$daily fluctuations. Energy storage can rapidly change its input$/$output power and shift demand over time, exhibiting great potential in supporting renewable power integration \cite{ESU}.

At the current stage, the unit capacity cost of energy storage is still relatively high, although it is continuously decreasing. The size of energy storage must be carefully determined. Existing works can be classified into two categories, according to the system scale. At the generation side, energy storage siting and sizing problem was studied in \cite{TN-1} and \cite{TN-2} via stochastic unit commitment and stochastic model predictive control under the multi-period economic dispatch framework. Joint capacity optimization of energy storage and transmission connector was discussed in \cite{TN-3} using a bi-level stochastic mixed-integer optimization in a market environment. In the above works, the uncertainty of renewable generation was represented by probability distributions and approximated through scenarios, or the operational risk was limited by chance constraints. Ref. \cite{TN-4} proposed two multi-parametric programming models to investigate the impact of energy storage on renewable spillage and flexibility enhancement. The optimal value function delivers useful information for storage sizing.   

At the distribution-level and demand side, ref. \cite{Small-1} proposed a reliability-constrained stochastic programming model for energy storage sizing in microgrids; supply inadequacy due to generator outage and intermittency of renewable plant was compensated by the energy storage unit. Storage sizing in island and grid-connect microgrids was discussed in \cite{Small-2}. The problem was revisited in \cite{Small-3} considering battery degradation, operating modes, and multiple choices of batteries. Ref. \cite{Small-4} presented a two-stage method for the optimal planning and operation of prosumer energy system. Storage was planned in the first stage aiming at minimizing life-cycle costs of renewable and storage facilities; the second-stage entailed a multi-objective energy management problem. Ref. \cite{Small-5} developed an optimization method to size the battery energy storage in electric vehicle parking lots. The uncertainty of charging demand was estimated by investigating the driving patterns and behaviors, such as the probability distributions of  arrival$/$departure time and driving distances.

This paper considers a particular scene: a large photovoltaic power station connects to the main grid via a long transmission corridor, as shown in Fig. \ref{fig:Renew-Line-System}. Given the long distance between the solar plant and the main grid, the unit capacity cost of the transmission line is usually much higher than that of the energy storage. Through a coordinated planning of line and storage capacities, the line capacity can be greatly reduced while maintaining renewable curtailment below a certain level. The contributions of this paper are twofold.

1) A data-driven robust formulation is established to evaluate the storage and transmission capacity   requirements of a remote solar plant. Distributional uncertainty of solar energy is captured by the perturbation of the probability coefficient associated with each day. The deterministic counterpart of the operation problem is derived based on duality theory.

2)  A linear programming based projection algorithm is developed to generate the feasible set of storage and transmission line capacity parameters. In the case study, we demonstrate how such a feasible set can help make an investment decision.

Unlike existing works which aim to provide a single planning strategy, the proposed method offers the entire feasible set of storage and transmission line capacities which ensures an efficient utilization of renewable energy. Such a method is useful in system planning and policy making, wherever long-distance transmission of renewable power is needed.

\section{Mathematical Model}

\begin{figure}[!t]
\centering
\includegraphics[scale=0.35]{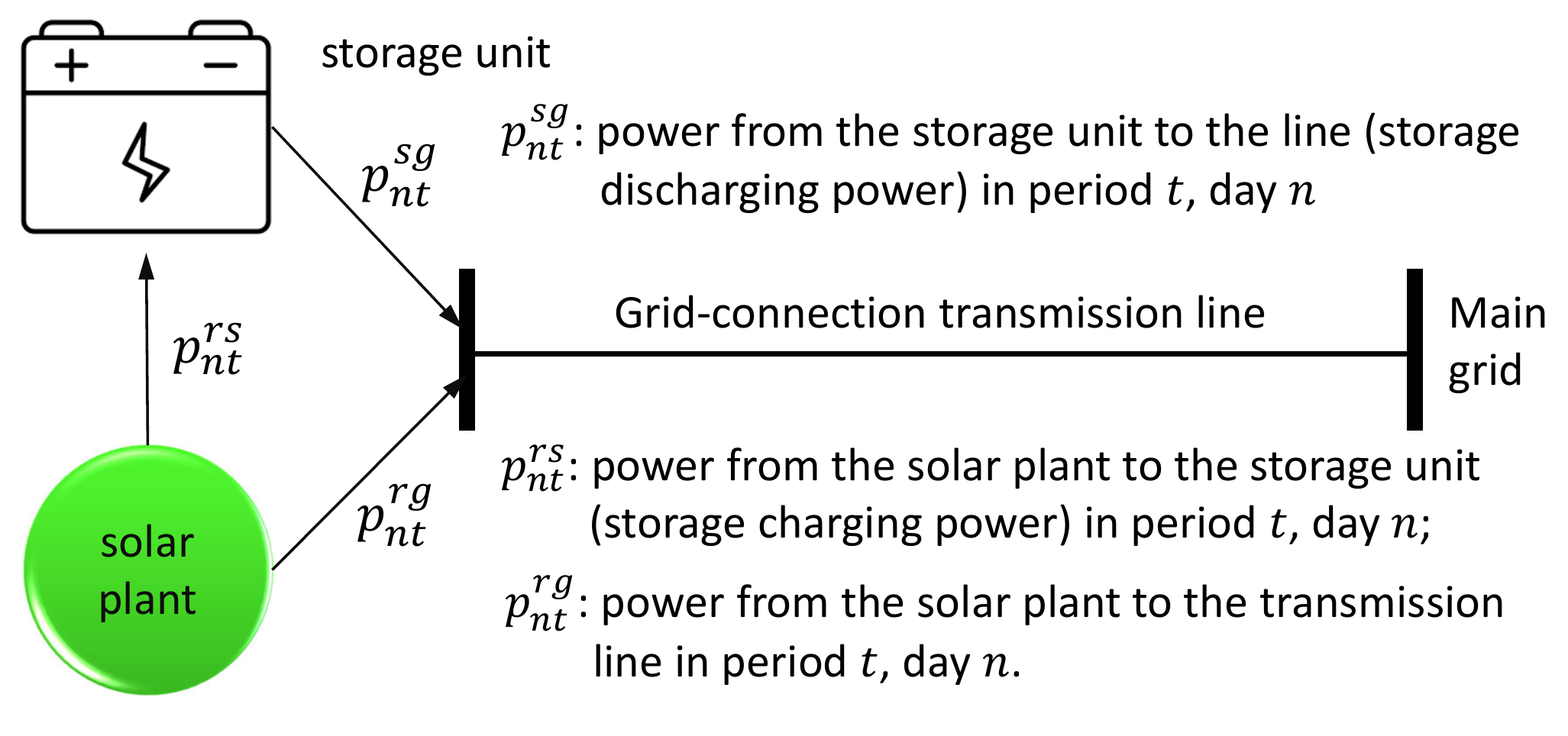}
\caption{Configuration of the remote solar generation system.}
\label{fig:Renew-Line-System}
\end{figure}

The remote solar power generation system in Fig. \ref{fig:Renew-Line-System} consists of a solar plant, an energy storage unit, and a transmission line. The components must interact and cooperate with each other to smooth the delivered power and achieve a lower renewable curtailment rate. The power flow relation is shown in Fig. 1.

\subsection{Energy Storage Model}

The storage model developed in \cite{ESU-Model} is used: 
\begin{subequations}
\label{eq:ESU-Operation}
\begin{equation} 
e_{t+1} = e_t + \eta^c p^c_t \Delta_t - {p^d_t \Delta_t}/{\eta^d} \label{eq:ESU-Oper-SOC}
\end{equation}
\begin{equation} 
\alpha_l e_m \le e_t \le \alpha_h e_m   \label{eq:ESU-Oper-SOC_Cap}
\end{equation}
\begin{equation}
p^c_t \ge 0,~ p^d_t \ge 0, p^c_t + p^d_t \le p_m  \label{eq:ESU-Oper-P-Cap}
\end{equation}
\end{subequations}
where $\eta^c/\eta^d$ represents the charging$/$discharging efficiency; $\Delta_t$ is the duration of period $t$; $p_m/e_m$ is the power$/$energy capacity of the storage unit, depending on the size of power electronics converter$/$battery array; $\alpha_l \in (0,0.5)$ and $\alpha_h \in (0.9,1)$ are constant coefficients implying storage operation limits. Decision variables include charging$/$discharging power $p^c_t/p^d_t$, as well as the state-of-charge (SoC) $e_t$. Constraints (\ref{eq:ESU-Oper-SOC})-(\ref{eq:ESU-Oper-SOC_Cap}) describe the dynamics and feasible range of SoC.

 In constraint (\ref{eq:ESU-Oper-P-Cap}), strict complementary condition $p^c_t \cdot p^d_t =0$ is relaxed. The storage unit can switch between charging and discharging, yielding more flexibility and positive $p^c_t$ and $p^d_t$ in the same period. Suppose $\Delta_t$ is divided into a charging interval $\Delta^c_t$ and a discharging interval $\Delta^d_t$, $p^c_t$ and $p^d_t$ are the average charging and discharging power. The physical charging power $\hat p_t^c$ and discharging power $\hat p_t^d$ can be calculated as 
\begin{equation}
\label{eq:ESU-RM-1}
\hat p_t^c = p^c_t \Delta_t/\Delta^c_t,~
\hat p_t^d = p^d_t \Delta_t/\Delta^d_t
\end{equation}
Physical implementation requires
\begin{equation}
\label{eq:ESU-RM-2}
\hat p_t^c \le p_m, \hat p_t^d \le p_m, \Delta^c_t+\Delta^d_t \le \Delta_t
\end{equation}
Substituting (\ref{eq:ESU-RM-1}) into (\ref{eq:ESU-RM-2}) gives $p_t^c \Delta t \le p_m \Delta_t^c$, $p_t^d \Delta t \le p_m \Delta_t^d$. Sum them up, we have $(p_t^c+p_t^d)\Delta t \le p_m (\Delta_t^c+\Delta_t^d) \le p_m \Delta t$, which is (\ref{eq:ESU-Oper-P-Cap}). Furthermore, for any feasible solution $(p^c_t,p^d_t)$ of model (\ref{eq:ESU-Operation}), the strategy $\hat p_t^c = \hat p_t^d =p_m$, $\Delta^c_t = p^c_t \Delta_t/p_m$,  $\Delta^d_t = p^d_t \Delta_t/p_m$ is always physically implementable. More details are available in\cite{ESU-Model}.

\subsection{Renewable Generation Model}

Historical data include hourly solar power output $p^r_{nt}$ in day $n=1:N$ and period  $t=1:24$ intraday. Each day is associated with an empirical probability $\rho^0_n$, $n=1:N$. Without loss of generality, we assume the empirical distribution is $\rho^0_1=\cdots=\rho^0_N=1/N$, which could be inexact, and the true probability $\rho=\left[\rho_1,\cdots,\rho_N\right]^\top $ resides in the following set    
\begin{equation}
\label{eq:Un-Set-1}
\Pi = \left\{ \rho \middle| \|\rho-\rho^0 \|_\infty \le \Gamma, 
\rho \ge 0, \|\rho\|_1 = 1   \right\}
\end{equation}
The first inequality restricts the distance between $\rho$ and $\rho^0$ by a constant $\Gamma$; the remaining constraints ensure $\rho$ is a valid probability distribution. If we expect the real distribution is contained in $\Pi$ with a confidence level of $\beta$, the recommended value of $\Gamma$ is \cite{Uncertainty}
\begin{equation}
\label{eq:Un-Set-2}
\Gamma = \dfrac{1}{2N} \ln \dfrac{2N}{1-\beta}
\end{equation}

\subsection{System Operation Model}

Following power flow variables defined in Fig. \ref{fig:Renew-Line-System}, the solar plant operation must obey the following constraints
\begin{subequations}
\label{eq:System}
\begin{equation}
\label{eq:System-1}
e_{nt+1} = e_{nt} + \eta^c p^{rs}_{nt} \Delta_t - {p^{sg}_{nt} \Delta_t}/{\eta^d},~
\forall n,~ \forall t
\end{equation}
\begin{equation}
\label{eq:System-2}
\alpha_l e_m \le e_{nt} \le \alpha_h e_m,~ \forall n,~ \forall t
\end{equation}
\begin{equation}
\label{eq:System-3}
p^{rs}_{nt} \ge 0, p^{sg}_{nt} \ge 0,~ p^{rs}_{nt} + p^{sg}_{nt} \le p_m,~\forall n,~ \forall t
\end{equation}
\begin{equation}
\label{eq:System-4}
p^{sg}_{nt} + p^{rg}_{nt} \le F_m,~p^{rg}_{nt} \ge 0, 
~\forall n,~ \forall t
\end{equation}
\begin{equation}
\label{eq:System-5}
p^{rg}_{nt} + p^{rs}_{nt} + \Delta p^r_{nt} =  p^r_{nt},
~\Delta p^r_{nt} \ge 0,~\forall n,~ \forall t
\end{equation}
\begin{equation}
\label{eq:System-6}
\sum_{n=1}^N \sum_{t=1}^{24} \rho_n  \Delta p^r_{nt} \le \sigma 
\sum_{n=1}^N \sum_{t=1}^{24} \rho_n  p^r_{nt},~ \forall \rho \in \Pi
\end{equation}
\end{subequations}
where (\ref{eq:System-1})-(\ref{eq:System-3}) are storage operation constraints; (\ref{eq:System-4}) limits the total power flow in the transmission line whose capacity is $F_m$; (\ref{eq:System-5}) prescribes power balancing, and the excessive power $\Delta p^r_{nt}$ is curtailed; the last inequality (\ref{eq:System-6}) imposes a cap $\sigma$ on renewable power spillage rate; In China, this value is $\sigma = 5\%$. To eliminate the enumeration of $\rho \in \Pi$ in (\ref{eq:System-6}), write it as
\begin{equation}
\label{eq:Deter-RC-P}
\begin{aligned}
\max_{\rho}~~& \sum_{n=1}^N \rho_n \left( \sum_{t=1}^{24} \left(\Delta p^r_{nt} - \sigma p^r_{nt} \right) \right) \le 0   \\
\mbox{s.t.}~~ & \rho \le 1 \cdot\Gamma + \rho^0: \mu^+,~
\rho \ge \rho^0 - 1 \cdot \Gamma : \mu^- \\
&  1^\top  \rho = 1 : \lambda, ~\rho \ge 0
\end{aligned}
\end{equation}
where the constraints interpret set $\Pi$; $\mu^+$, $\mu^-$, and $\lambda$ following a colon is the dual variable associated with each constraint. 

Because strong duality holds for feasible linear programs, condition (\ref{eq:Deter-RC-P}) requires that the optimum of the dual objective should be non-positive, i.e.:
\begin{equation}
\label{eq:Deter-RC}
\begin{gathered}
{\bf 1}^\top (\mu^+ - \mu^-) \Gamma + (\mu^+ + \mu^-)^\top \rho^0 + \lambda \le 0,~ \mu^+  \ge 0 \\
\mu^+_n + \mu^-_n + \lambda  \ge  \sum\nolimits_{t=1}^{24} \left(\Delta p^r_{nt} - \sigma p^r_{nt} \right), \forall n,~ \mu^- \le 0 
\end{gathered}
\end{equation}

\begin{figure*}[!t]
\centering
\includegraphics[scale=0.6]{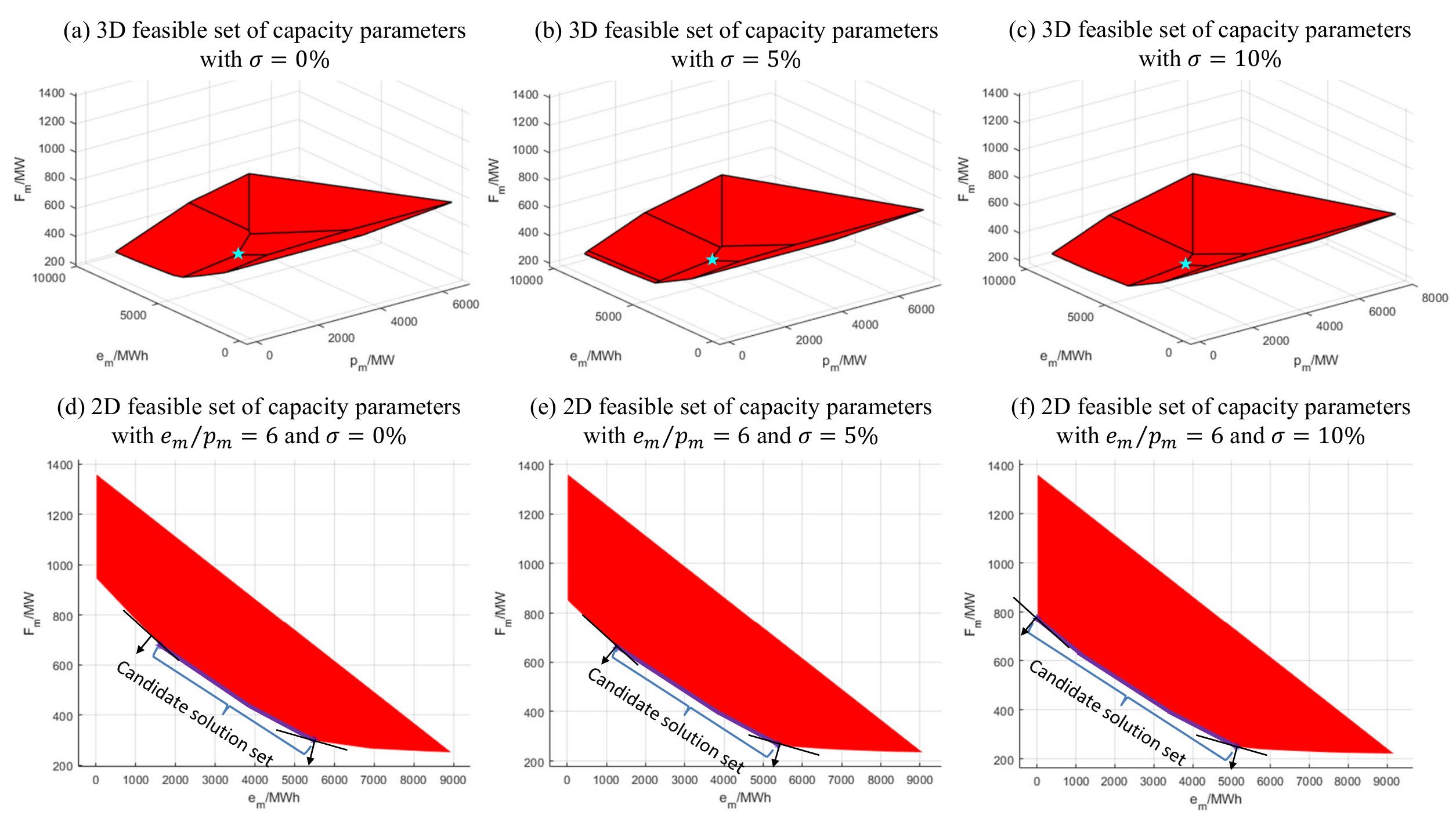}
\caption{Feasible set of capacity parameters under different caps.}
\label{fig:Solar-PEF-Set}
\end{figure*}

\section{Feasible Set of Capacity Parameters}

For notation brevity, we use a compact form. Vector $x$ includes all dispatch variables $p^{rs}_{nt}$, $p^{rg}_{nt}$, $p^{sg}_{nt}$, $e_{nt}$, $\Delta p^r_{nt}$, $\forall n, \forall t$ and dual variables $\mu^+$, $\mu^-$, $\lambda$. Vector $\theta=[p_m,e_m,F_m]^\top$ encompasses capacity parameters. The operation problem entails finding a feasible solution in
\begin{equation}
\label{eq:Compact-1}
\Lambda(\theta) = \{x ~|~ (\ref{eq:System-1})-(\ref{eq:System-5}),(\ref{eq:Deter-RC})\}
\end{equation}
under a fixed $\theta$, where all constraints are linear. So (\ref{eq:Compact-1}) can be written in a compact matrix form as
\begin{equation}
\label{eq:Compact-2}
\Lambda(\theta) = \{x ~|~ Ax + B\theta \le b\}
\end{equation}
In (\ref{eq:Compact-2}), $\theta$ is regarded as a parameter, and equality constraints are equivalently expressed via two opposite inequalities. 

Suppose the unit capacity cost of power electronics converters, battery array, and transmission line is $c_p$, $c_e$, and $c_l$, respectively; the available investment budget is $\xi_m$; vector $c=[c_p,c_e,c_l]^\top$. The feasible region of $\theta$ is defined as 
\begin{equation}
\label{eq:Compact-Theta}
\Theta = \{\theta ~|~ \Lambda(\theta) \ne \emptyset, c^\top \theta \le \xi_m\}
\end{equation}
The non-empty requirement inspires a projection formulation. 

Define a polyhedron in variables $x$ and $\theta$ 
\begin{equation}
\label{eq:Compact-P}
\mathbb P = \{(x,\theta) ~|~ Ax + B\theta \le b, c^\top \theta \le \xi_m\}
\end{equation}
Then $\Theta$ is the projection of $\mathbb P$ onto the subspace spanned by the coordinates of $\theta$. According to the projection theorem in \cite{Projection}, $\Theta$ can be expressed as
\begin{equation}
\label{eq:Theta-Closed-Thm}
\Theta = \{\theta~|~ \gamma^\top B \theta \ge \gamma^\top b,\forall \gamma \in {\rm{vert}}(D), c^\top \theta \le \xi_m\}
\end{equation}
where $D = \{\gamma~|~ A^\top \gamma = 0, -1 \le \gamma \le 0\}$, and vert$(D)$ denotes all the vertices of $D$. However, vertex enumeration in (\ref{eq:Theta-Closed-Thm}) is not a practical method, because the dimension of $D$ is high.

By  (\ref{eq:Theta-Closed-Thm}) we have
\begin{equation}
\label{eq:Alg-1}
\gamma^\top B \theta \ge \gamma^\top b,~ 
\forall \gamma \in D,~ \forall \theta \in \Theta
\end{equation}
which indicates that if $\theta^* \notin \Theta$, there must be some $\gamma^* \in D$ satisfying $(\gamma^*)^\top (b - B \theta^*) > 0$. Therefore, the hyperplane
\begin{equation}
\label{eq:Alg-3}
(\gamma^*)^T B \theta = (\gamma^*)^\top b 
\end{equation}
strictly separates $\theta^*$ from $\Theta$. As (\ref{eq:Alg-3}) will not remove any interior point in $\Theta$ which satisfies $(\gamma^*)^\top (b - B \theta^*) < 0$, (\ref{eq:Alg-3}) is the boundary of $\Theta$. 

The  strategy for computing $\Theta$ is to create a large enough initial set $\Theta_{temp}$ which contains $\Theta$. Then remove $\theta \notin \Theta$ by (\ref{eq:Alg-3}), until (\ref{eq:Alg-1}) is met. This entails solving
\begin{equation}
\label{eq:BLP}
\begin{aligned}
v^* = \max~~ & \gamma^\top (b - B \theta)  \\ 
\mbox{s.t.}~~& \gamma \in D,~ \theta \in \Theta_{temp}  
\end{aligned}
\end{equation}
Since $0 \in D$ and $\gamma=0$ is feasible, the optimum $v^*$ must be non-negative. If $v^*=0$, then (\ref{eq:Alg-1}) is certified; otherwise, if $v^* > 0$, equality (\ref{eq:Alg-3}) with the optimal solution $\gamma^*$ generates a boundary facet of $\Theta$. However, bilinear program (\ref{eq:BLP}) is non-convex; a local optimal solution is insufficient to  certify (\ref{eq:Alg-1}). 

It is proven that the optimal solution of a bilinear program like (\ref{eq:BLP}) must be found at the vertices of $D$ and $\Theta_{temp}$. As $\theta \in \mathbb R^3$, the dimension of $\Theta_{temp}$ is low, we are able to enumerate its vertices as $\mbox{vert}(\Theta_{temp}) = \{\theta_k\}_{k=1}^K$. Then, we solve $K$ linear programs as follows 
\begin{equation}
\label{eq:BLP-LPs-1}
v^*_k = \max_{\gamma \in D} \gamma^\top (b - B \theta_k)
\end{equation}
The optimal solution and optimal value of the $k$-th problem are $\gamma^*_k$ and $v^*_k$, respectively. The maximum of the $K$ optimums is the global optimum of bilinear program (\ref{eq:BLP}), i.e.:
\begin{equation}
\label{eq:BLP-LPs-2}
\{v^*,k^*\} : \max \{v^*_k\},~ \gamma^* = \gamma^*_{k^*}
\end{equation}
The flowchart of the linear programming based projection method is summarized in Algorithm 1.

\begin{algorithm}[!t]
\small
\caption{{\bf  }}
\begin{algorithmic}[1]
\State Initiation: $\Theta_{temp}= \{\theta | \theta \ge 0,~ c^\top \theta \le \xi_m\}$. 

\State Update vert$(\Theta_{temp})$; find unvisited vertices.
 
\State Solve problem (\ref{eq:BLP-LPs-1}) corresponding to unvisited vertices.
\State Update $v^*$ and $\gamma^*$ by (\ref{eq:BLP-LPs-2}) If $v^* = 0$, terminate; if $v^*>0$, add a cut $(\gamma^*)^\top B \theta \ge (\gamma^*)^\top b$ in $\Theta_{temp}$, and go to step 2.  
\end{algorithmic}
\label{Alg:Cutting-Plane}
\end{algorithm}

\section{Case Studies}

\begin{figure}
\begin{minipage}[c]{.5\linewidth}
\renewcommand{\arraystretch}{1.8}
\tiny
\makeatletter\def\@captype{table}\makeatother 
\label{fig:Strategy}
\caption{Optimal sizing strategies}
\setlength{\tabcolsep}{2mm}{\begin{tabular}{c|ccc}
\toprule
Spillage rate $\sigma $      &  $0\%$  &  $5\%$  &   $10\%$  \\  
\midrule
Iteration   &       42       &      38        &        32    \\
Time (sec.) &      38.3      &     34.2       &      28.8    \\
$p_m$(MW)   &      256.1     &     152.6      &      130.4   \\ 
$e_m$(MWh)  &      1239      &     1009       &      805.7   \\
$F_m$(MW)   &      704.5     &     680.0      &      648.2   \\
cost ($10^9$CNY)&  9.492     &     8.844      &      8.227   \\
\bottomrule
\end{tabular}}
\end{minipage}%
\begin{minipage}[c]{.5\linewidth}
\centering
\includegraphics[scale=0.33]{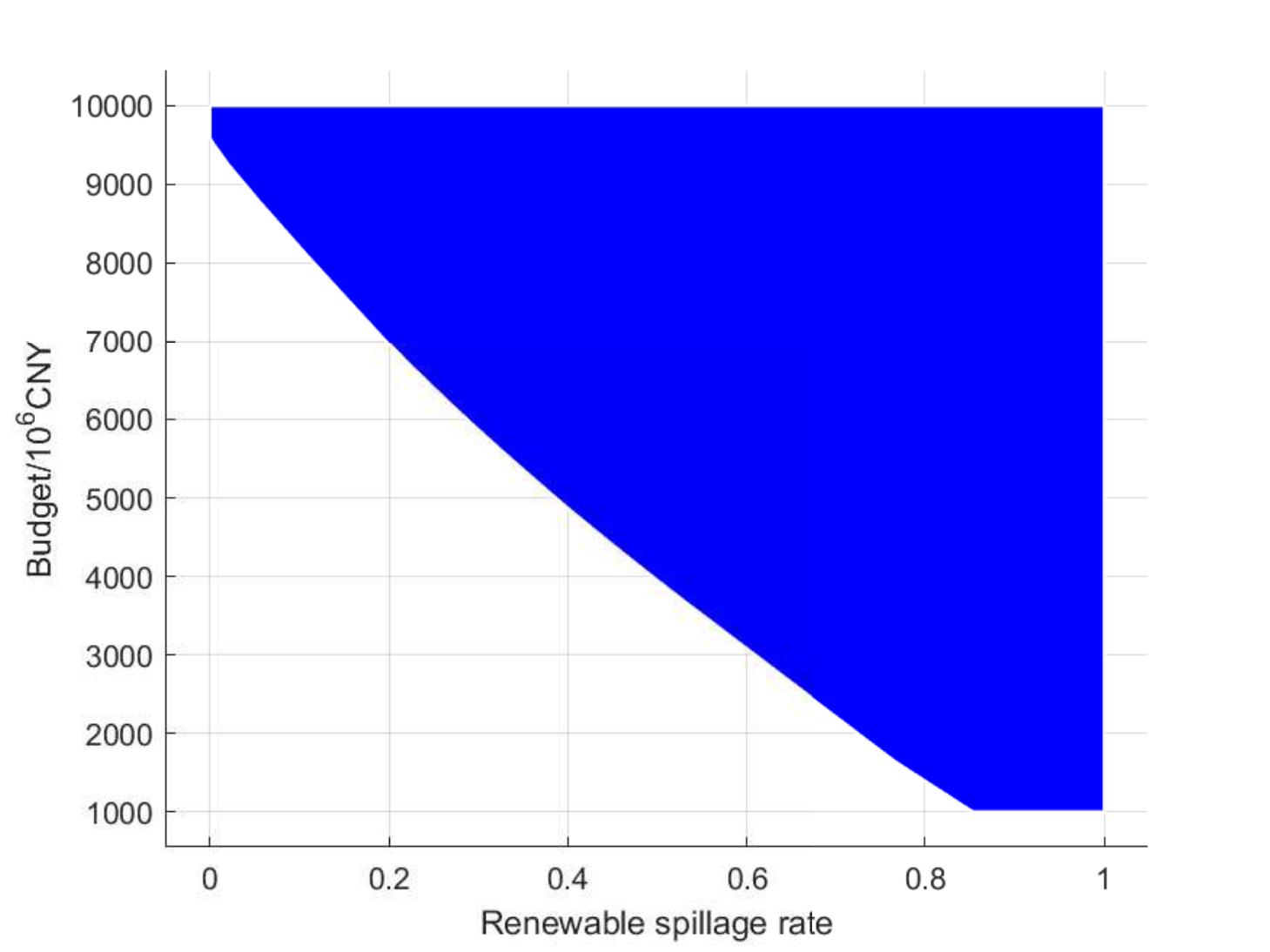}
\caption{Relation between $\sigma$ and $\xi_m$.}
\label{fig:sigma-budget}
\end{minipage}
\end{figure}

The hourly solar radiation data recorded  in the south of Qinghai Province, China during 2019 are used. We choose the data in 120 days, 10 in each month,  to build the output curve of an 1GW solar power station in planning. In the uncertainty model, $\beta$ is set to $0.99$. According to (\ref{eq:Un-Set-2}), $\Gamma=0.042$ is used, implying that in the worst-case distribution, a day can have a maximal probability of $1/N+\Gamma$ $=5.03\%$. For the energy storage, $\eta^c = \eta^d = 0.95$, $\alpha_l=0.25$ and $\alpha_h=0.95$. The unit capacity cost coefficients are $c_p=10^6$\yen$/$MW, $c_e=1.2\times 10^6$ \yen$/$MWh, and $c_l=1.1\times 10^7$\yen$/$MW; $\xi_m = 1.5\times 10^{10}$\yen.

The feasible sets of capacity parameters $(p_m,e_m,F_m)$ with $\sigma=0,5\%, 10\%$ are plotted in Figs. \ref{fig:Solar-PEF-Set}a-\ref{fig:Solar-PEF-Set}c. With the increase of $\sigma$, more renewable power spillage is allowed, leading to a larger feasible set. If we solve $\min_{\theta \in \Theta} c^\top \theta$, the optimal solution provides the sizing strategy with the minimal investment cost. Results are listed in Table I. In the three cases, Algorithm~1 converges in about 40 iterations; the computation time is about half a minute. Hence, the computational efficiency is satisfactory. With the help of energy storage, the capacity of transmission line is about $68\%$ of the capacity of solar station when $\sigma=5\%$ (stipulated in China), which greatly reduces the burden on transmission line construction. The optimal energy-power capacity ratio is about $5\sim7$ hour in all the three cases, which mainly depends on the solar output curve. The cap of renewable spillage rate has notable influence on the size of storage unit which is much cheaper, and its impact on the transmission line  capacity is not so significant.   

The impact of storage cost can be analyzed based on the feasible sets. According to above results, if we fix $e_m/p_m=6$, then the feasible sets in $\mathbb R^2$ are portrayed in Figs. \ref{fig:Solar-PEF-Set}d-\ref{fig:Solar-PEF-Set}f. The difference is clear: without energy storage ($e_m=0$), the feasible capacity of the transmission line decreases with the growth of $\sigma$. Since the energy-power ratio is given, the unit capacity cost of energy storage is $c^0_s =c_e+c_p/6\approx 1.367 \times 10^6$ \yen$/$MWh. We change $c_s$ in the interval $[0.5c^0_s,1.2 c^0_s]$. Coefficients $c_s$ and $c_l$ determine the gradient vector of the investment cost, and thus influence the optimal sizing strategy. Accounting for the continuous variation of $c_s$, the gradient vector of investment cost is illustrated in Fig. \ref{fig:Solar-PEF-Set}d. The optimal sizing strategy is one of the vertices of the feasible region determined by the gradient vector, as in Fig. \ref{fig:Solar-PEF-Set}d. In the same way, all candidate solutions in each case are marked in Figs. \ref{fig:Solar-PEF-Set}d-\ref{fig:Solar-PEF-Set}f. When $\sigma=10\%$ and the energy storage is expensive, only transmission line is invested. In all the remaining cases, energy storage plays an important role in reducing the line capacity and the total investment cost. 

The feasible set in Fig. 2 offers more insightful information. The Pareto front of the feasible set consists of the points where capacities of energy storage $e_m$ and transmission line $F_m$ cannot be reduced simultaneously. The Pareto front does not depend on the costs of storage and transmission line. From Fig. 2 we can see that if $e_m/p_m=6$ hour, at the Pareto front, reducing line capacity by 100MW requires the deployment of about 1GWh energy storage. Cost information is needed only when a concrete optimal solution is needed. If the cost function is nonlinear, the optimal solution can be observed by plotting the contour of the objective function. The proposed method simplifies the operation problem which involves much more variables and retains the operational requirements on the capacities, making capacity optimization quite straightforward.

If the available budget $\xi_m$ shrinks, the facet corresponding to the budget constraint in Figs. \ref{fig:Solar-PEF-Set}a-\ref{fig:Solar-PEF-Set}f moves towards the origin; the feasible set may become a singleton, which determines the optimal sizing strategy under the given spillage cap. We investigate the relationship between the budget and the corresponding minimum spillage rate. By treating $\xi_m$ and $\sigma$ in (\ref{eq:System-1})-(\ref{eq:System-5}), (\ref{eq:Deter-RC}) and (\ref{eq:Compact-P}) as variables, we can project polyhedron $\mathbb P$ onto the subspace spanned by the coordinates of $\xi_m$ and $\sigma$. In this case, capacities $p_m,e_m,F_m$ are also decision variables;  the parameter is $\theta=[\sigma,\xi_m]^\top$, and the parameter set is $\Theta = \{(\sigma,\xi_m)|\sigma \in [0,1], \xi_m \in [10^9, 10^{10}] \}$. Executing Algorithm 1, the relation between $\xi_m$ and $\sigma$ is obtained and depicted in Fig. \ref{fig:sigma-budget}. This figure clearly shows how the spillage rate influences the minimum budget. If $\sigma=0$, the minimum budget is $\xi_m=9.5\times10^6$\yen. Such an illustrative result provides useful information for capacity sizing and policy making.

\section{Conclusions}

This paper studies the capacity requirements of storage and transmission line in order to achieve a certain spillage target for a remote solar generation system. A linear programming based projection algorithm is proposed to determine the feasible set of capacity parameters.  The proposed method can offer useful information for capacity sizing and policy making.

\ifCLASSOPTIONcaptionsoff
  \newpage
\fi

\bibliographystyle{IEEEtran}
\bibliography{IEEEabrv,refs}

\end{document}